\documentclass[3p,times,twocolumn]{elsarticle}
 \biboptions{comma,sort&compress}

\usepackage{graphicx}
\usepackage{subfigure}
\usepackage{slashed}
\usepackage{multirow}
\usepackage{amsmath}
\usepackage{here}
\usepackage{ecrc}

\volume{00}

\firstpage{1}

\journalname{Nuclear and Particle Physics Proceedings}
\runauth{Niu Su}

\jid{nppp}
\jnltitlelogo{Nuclear and Particle Physics Proceedings}

\usepackage{amssymb}

\usepackage[figuresright]{rotating}

\begin{document}

\begin{frontmatter}

\title{
%
Investigation on the $\Omega(2012)$ from QCD sum rules\,$^*$}

 \cortext[cor0]{Talk presented at QCD24, 27th International Conference in QCD (8-12/07/2024,
  Montpellier - FR). }

 \author[label1,label2]{Niu Su}
 \ead{suniu@seu.edu.cn}
 \author[label2]{Hua-Xing Chen}
 \author[label3]{Philipp Gubler}
 \author[label1,label3]{Atsushi Hosaka}
\corref{cor1}
   \address[label1]{Research Center for Nuclear Physics (RCNP), Osaka University, Ibaraki 567-0047, Japan
}
\address[label2]{
School of Physics, Southeast University, Nanjing 210094, China
}
\address[label3]{
Advanced Science Research Center, Japan Atomic Energy Agency (JAEA), Tokai 319-1195, Japan
}
\cortext[cor1]{Speaker}

\pagestyle{myheadings}
\markright{ }
\begin{abstract}
\noindent
We investigate the recently observed $\Omega(2012)$ baryon using QCD sum rules. By constructing $P$-wave $\Omega$ baryon currents and performing spin projection and parity projection, we obtain the masses of the $J^P = 1/2^-$ and $3/2^-$ states as $M_{1/2^-} = 2.07^{+0.07}_{-0.07}{\rm~GeV}$ and $M_{3/2^-} = 2.05^{+0.09}_{-0.10}{\rm~GeV}$, in good agreement with experiment. This suggests that $\Omega(2012)$ is likely to be a negative parity $P$-wave excited state, though its spin remains undetermined and requires further study of its decay properties.

\end{abstract}
\end{frontmatter}
\section{Introduction}

In 2018, the exited $\Omega$ baryon, $\Omega(2012)$, was observed for the first time in $\Upsilon(1S), \Upsilon(2S)$, and $\Upsilon(3S)$ decays by the Belle experiment~\cite{Belle:2018mqs}.The experimental evidence has been further strengthened by the $\Omega_c$ decay~\cite{Belle:2021gtf}.
The latest data for its mass and decay width are~\cite{Belle:2022mrg}:
\begin{eqnarray}
M &=& 2012.5 \pm 0.7 \pm 0.5 {\rm~MeV} \, ,
\\   \nonumber
\Gamma &=& 6.4^{+2.5}_{-2.0}  {\rm~MeV} \, .
\end{eqnarray}

The conventional quark model may naively explain the $\Omega(2012)$ to be a negative parity state
as the first $P$-wave excitation of the ground-state $\Omega$ baryon with three strange quarks~\cite{Aliev:2018syi,Aliev:2018yjo,Polyakov:2018mow,Wang:2018hmi,Xiao:2018pwe,Liu:2019wdr,Arifi:2022ntc,Menapara:2021vug,Wang:2022zja,Zhong:2022cjx}.
One important feature of this quark model picture is that there should be spin-orbit partners of both $J^P = 1/2^-$ and $3/2^-$. In contrast, because of the fact that the mass of $\Omega(2012)$ is close to the
$\bar K$ and $\Xi^*(1530)$ threshold, a molecular picture of these particles
has been proposed and extensively discussed in Refs.~\cite{Lin:2018nqd,Valderrama:2018bmv,Pavao:2018xub,Huang:2018wth,Gutsche:2019eoh,Ikeno:2020vqv,Zeng:2020och,Lu:2020ste,Liu:2020yen,Ikeno:2022jpe,Hu:2022pae}. In this situation, a hybrid picture of three-quark and molecular structures was proposed~\cite{Lu:2022puv}. After all, at this moment it would be fair to say that the structure of $\Omega(2012)$ is not yet well understood, and this has motivated us to study further properties of this state in yet another theoretical approach based on QCD sum rules.

\section{$P$-wave $\Omega$ baryon currents}

Let us first construct the currents for the $P$-wave $\Omega$ baryon using three strange quark fields $s_a(x)$. We follow a process similar to the Ioffe argument~\cite{Ioffe:1981kw}, incorporating a non-vanishing diquark structure. When a derivative is present, it can be shown that three types of diquarks do not vanish:
$\epsilon^{abc} s_a^{T} C \gamma_5 {\overset{\leftrightarrow}{D}}_\mu s_b $, $\epsilon^{abc} s_a^{T} C {\overset{\leftrightarrow}{D}}_\mu s_b$ and
$\epsilon^{abc} s_a^{T} C \gamma_\mu \gamma_5 {\overset{\leftrightarrow}{D}}_\mu s_b$, which are the scalar, vector, and axial-vector types, respectively.
Here $D_\mu = \partial_\mu + i g_s A_\mu$ is the covariant derivative with the gluon field $A_\mu$.

In this work we employ the following $ss$-diquark
\begin{equation}
\epsilon^{abc} [s_a^T C \gamma_5 {\overset{\leftrightarrow}{D}}_\mu s_b] = - 2 \epsilon^{abc} [(D_\mu s_a^T) C \gamma_5 s_b] \,
\label{eq_diquark_Our_Choice},
\end{equation}
This diquark has a more appropriate internal $P$-wave structure compared to the other two diquarks. By combining this diquark with the third quark field, which has a spin of $1/2$, one can construct the currents for the $P$-wave $\Omega$ baryon with a total angular momentum of $J_{tot} = 1/2$ or $3/2$ as follows
\begin{eqnarray}
J &=& -2\epsilon^{abc} ~ [(D^\mu s_a^T) C \gamma_5 s_b]
\label{def:current1}~ \gamma_\mu s_c \, ,
\\
J_\mu &=& -2\epsilon^{abc} ~ [(D^\nu s_a^T) C \gamma_5 s_b] \label{def:current2}(g_{\mu\nu} -
{1\over4}\gamma_\mu\gamma_\nu) s_c \, .
\end{eqnarray}
We will employ them in the present study.

\section{QCD sum rule analyses}

As an example, the current $J_\mu$, which has spin and parity $J^P = 3/2^-$, can couple to the physical state $|\Omega; 3/2^-\rangle$ with the corresponding matrix element expressed as follows:
\begin{eqnarray}
\langle 0 | J_\mu | \Omega; 3/2^- \rangle = f_{-} u_\mu(q)\, ,
\label{coupling1}
\end{eqnarray}
where
$f_-$ is a coupling  constant and $u_\mu(q)$ the Rarita-Schwinger vector-spinor. The current $J_\mu$ can also couple to a positive parity state $|\Omega; 3/2^+ \rangle$ with the matrix element given as follows:
\begin{eqnarray}
\langle 0 | J_\mu | \Omega;3/2^+ \rangle = f_{+} \gamma_5 u_\mu(q)\, .
\label{coupling2}
\end{eqnarray}

We next study the correlation function with the following Lorentz structure:
\begin{eqnarray}
\Pi_{\mu\nu}(q^2) &=& i \int d^4x e^{iqx} \langle 0 | {\bf T}[J_{\mu}(x) J_\nu^\dagger(0)] | 0 \rangle \label{twoponit}
\\ \nonumber
&=&(\frac{q_\mu q_\nu}{q^2}-g_{\mu\nu})\Pi(q^2)+\cdots \, .
\end{eqnarray}
$\Pi(q^2)$ can also be expressed as a dispersion relation,
\begin{eqnarray}
\Pi(q^2) = \int^\infty_{s_<}\frac{\rho(s)}{s-q^2-i\varepsilon}ds \, ,
\end{eqnarray}
where $\rho(s) \equiv {\rm Im} \Pi(s)/\pi$ is the spectral density, and $s_< = 9 m_s^2$ is the lower threshold of the spectral function computed by the OPE.

At the hadron level, the spectral density is obtained by inserting the complete set of intermediate hadronic states
\begin{eqnarray}
\rho^{\rm phen}(s)  &\equiv& \sum_n\delta(s-M^2_n) \langle 0| J_{\mu} | n\rangle \langle n| J_{\nu}^{\dagger} |0 \rangle \label{eq:rho}
\\ \nonumber
&=&f_-^2 (\slashed{q}+M_-) \delta(s-M_-^2)
\\ \nonumber
&&+ f_+^2 (\slashed{q}-M_+) \delta(s-M_+^2)
\\ \nonumber
&&+ \theta(s-s_0)\rho^{\rm cont}(s) \, ,
\end{eqnarray}
in which we consider two poles for both $|\Omega; 3/2^-\rangle$ and $|\Omega; 3/2^+\rangle$ as well as the continuum contribution.
Based on Eqs.~(\ref{coupling1})-(\ref{eq:rho}), the correlation function can be given as
\begin{eqnarray}
\nonumber \Pi^{\rm phen}(q^2)&\!\!=\!\!&f_-^2 \frac{\slashed{q}+M_-}{M_-^2-q^2 - i \epsilon}+ f_+^2 \frac{\slashed{q}-M_+ }{M_+^2-q^2 - i \epsilon}
\\ &\!\!=\!\!& \Pi_1^{\rm phen}(q^2)\slashed{q}+\Pi_0^{\rm phen}(q^2)\, ,
\end{eqnarray}
By introducing the spectral densities $\rho^{\rm phen}_{0,1}$ for $\Pi^{\rm phen}_{0,1}$, we can write down the following relations
\begin{eqnarray}
\rho^{\rm phen}_1(s)\!\!\!\!&=&\!\!\!\!f_-^2\delta(s-M^2_-)+ f_+^2\delta(s-M^2_+)  \label{phen1}\, ,
\\\rho^{\rm phen}_0(s)\!\!\!\!&=&\!\!\!\!f_-^2M_-\delta(s-M^2_-)- f_+^2M_+\delta(s-M^2_+) \label{phen0}\, ,
\end{eqnarray}
from which we can extract the spectral densities for negative and positive parity states as
\begin{eqnarray}
\rho^{\rm phen}_{\mp}(s) = \sqrt{s} \rho^{\rm phen}_1(s)\pm \rho^{\rm phen}_0(s) .
\end{eqnarray}

At the quark-gluon level, we use the method of the operator product expansion to calculate the correlation function in Eq.~(\ref{twoponit}),
from which we extract the spectral densities $\rho^{\rm OPE}_1(s)$ and $\rho^{\rm OPE}_0(s)$.

By equating the spectral densities at the
hadron and quark-gluon levels and
applying the Borel transformation, we derive the sum rules as
\begin{eqnarray}
\Pi_\mp(s_0,M_B)&=& 2M_{\mp}f_{\mp}^2 e^{-M^{2}_{\mp}/M_B^2} \label{sumrule}
\\ \nonumber &=& \int^{s_0}_{s_<} (\sqrt{s} \rho^{\rm OPE}_1(s) \pm \rho^{\rm OPE}_0(s))e^{-s/M_B^2}ds .
\end{eqnarray}

The masses and coupling constants are obtained by the formulae
\begin{eqnarray}
\label{eq:mass} && M^2_{\mp}(s_0, M_B)
\\ \nonumber&=& \frac{\int^{s_0}_{s_<} (\sqrt{s} \rho^{\rm OPE}_1(s) \pm \rho^{\rm OPE}_0(s)) s e^{-s/M_B^2} ds}{\int^{s_0}_{s_<} (\sqrt{s} \rho^{\rm OPE}_1(s) \pm \rho^{\rm OPE}_0(s)) e^{-s/M_B^2} ds} ,
\end{eqnarray}
and
\begin{eqnarray}
\label{eq:decay} &&f^2_{\mp}(s_0, M_B)
\\\nonumber&=&\frac{\int^{s_0}_{s_<} (\sqrt{s} \rho^{\rm OPE}_1(s) \pm \rho^{\rm OPE}_0(s)) e^{-s/M_B^2} ds \times e^{M_{\mp}^2/M_B^2}}{2M_{\mp}} .
\end{eqnarray}
\section{Numerical analysis}

The used input
parameters for condensates and masses determined at
the renormalization scale 2~GeV are~\cite{pdg,Ovchinnikov:1988gk,Yang:1993bp,Ellis:1996xc,Ioffe:2002be,Jamin:2002ev,Gimenez:2005nt,Narison:2011xe,Narison:2018dcr}:
\begin{eqnarray}
\langle\bar qq \rangle &=& -(0.240 \pm 0.010)^3 \mbox{ GeV}^3 \, ,\label{eq:condensate}
\\ \nonumber \langle\bar ss \rangle &=& (0.8\pm 0.1)\times \langle\bar qq \rangle \, ,
\\ \nonumber \langle g_s\bar s\sigma G s\rangle &=&  (0.8 \pm 0.2)\times\langle\bar ss\rangle \, ,
\\ \nonumber \langle \alpha_s GG\rangle &=& (6.35 \pm 0.35) \times 10^{-2} \mbox{ GeV}^4 \, ,
\\ \nonumber m_s &=& 93 ^{+9}_{-3} \mbox{ MeV} \, .
\end{eqnarray}

Let us discuss the $J^P = 3/2^-$ state as an example. From Eq.~(\ref{eq:mass}), we see that the mass depends on two free parameters: the Borel mass $M_B$ and the threshold value $s_0$. To determine the appropriate working regions for these parameters, we use the following three criteria: a) sufficiently good convergence of the OPE,
b) sufficiently large pole contribution, and c) sufficiently weak mass dependence on these two parameters.

\begin{figure}[hbt]
\begin{center}
\includegraphics[width=0.47\textwidth]{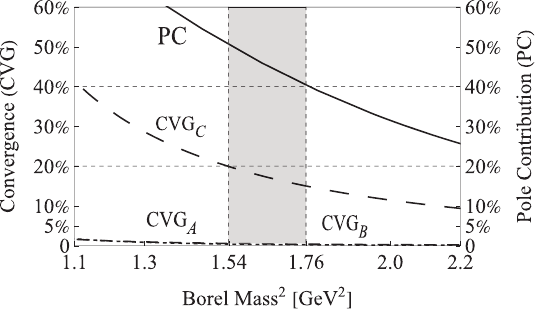}
\caption{CVG$_{A/B/C}$ and PC as functions of the Borel mass $M_B$ when $s_0$ is set to 6.0~GeV$^2$. CVG$_A$ (short-dashed) and CVG$_B$ (medium-dashed) are almost overlaid.}
\label{fig:cvgpole}
\end{center}
\end{figure}

To ensure the OPE convergence, we require that
\begin{eqnarray}
\mbox{CVG}_A &\equiv& \left|\frac{ \Pi_-^{{\rm D=11+10+9+8}}(\infty, M_B^2) }{ \Pi_-(\infty, M_B^2) }\right| \leq 5\% \, ,
\\ 
\mbox{CVG}_B &\equiv& \left|\frac{ \Pi_-^{{\rm D=7+6}}(\infty, M_B^2) }{ \Pi_-(\infty, M_B^2) }\right| \leq 10\% \, ,
\\ 
\mbox{CVG}_C &\equiv& \left|\frac{ \Pi_-^{{\rm D=5+4}}(\infty, M_B^2) }{ \Pi_-(\infty, M_B^2) }\right| \leq 20\% \, ,
\label{eq:CVG_C}
\end{eqnarray}
As shown in Fig.~\ref{fig:cvgpole} with the three dashed curves, we find that the Borel mass $M_B^2$ must be larger
than 1.54 GeV$^2$.

For a sufficient pole contribution we require
\begin{equation}
\mbox{PC} \equiv \left|\frac{ \Pi_-(s_0, M_B^2) }{ \Pi_-(\infty, M_B^2) }\right| \geq 40\% \, .
\label{eq:pole_contribution_cond}
\end{equation}
As shown in Fig.~\ref{fig:cvgpole} with the solid curve, we find that the Borel mass $M_B^2$ must be less than 1.76 GeV$^2$ when $s_0 = 6.0$~GeV$^2$. In the analysis of the above two criteria, we noticed that $s_0$ has a minimum value $s^{\rm min}_0 = 5.3$~GeV$^2$, and we have chosen $s_0$ slightly larger than it. Altogether the Borel window is determined to be $1.54$~GeV$^2 \leq M_B^2 \leq 1.76$~GeV$^2$ when $s_0 = 6.0$~GeV$^2$, and the working region of $s_0$ is determined to be $5.3$~GeV$^2 \leq s_0 \leq 6.7$~GeV$^2$.

\begin{figure*}[hbtp]
\begin{center}
\subfigure[]{\includegraphics[width=0.35\textwidth]{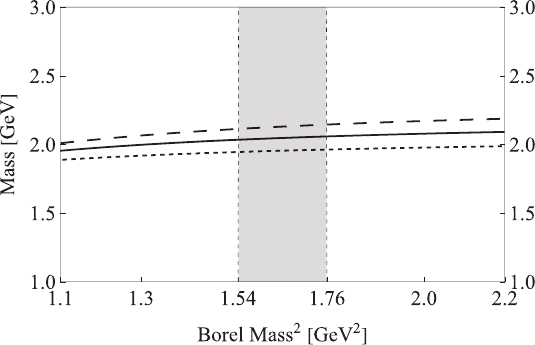}}
~~~~~~~~~~
\subfigure[]{\includegraphics[width=0.35\textwidth]{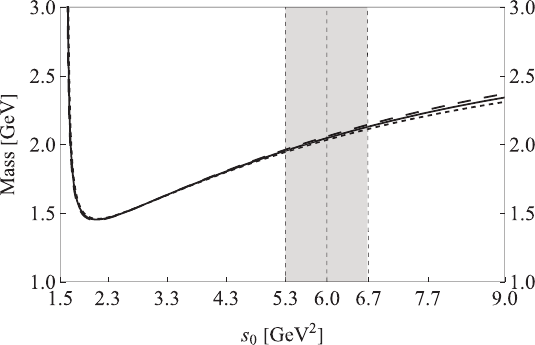}}
\caption{The mass $M_{3/2^-}$ as a function of the Borel mass $M_B^2$ and the threshold value $s_0$ extracted from the current $J_{\mu}$ in Eq~(\ref{def:current2}). In the left panel, the short-dashed/solid/long-dashed curves are obtained by setting $s_0 = 5.3/6.0/6.7$~GeV$^2$, respectively. In the right panel, the short-dashed/solid/long-dashed curves are obtained by setting $M_B^2 = 1.54/1.65/1.76$~GeV$^2$, respectively.}
\label{fig:mass}
\end{center}
\end{figure*}

We can now study the mass of the $3/2^-$  state as a function of the Borel mass $M_B^2$ and the threshold value $s_0$ as shown in Fig.~\ref{fig:mass}. From the left panel, we see that the mass is nearly independent of $M_B^2$ within the region $1.54$~GeV$^2 \leq M_B^2 \leq 1.76$~GeV$^2$. From the right panel, the mass dependence on $s_0$ is acceptable in the region $5.3$~GeV$^2 \leq s_0 \leq 6.7$~GeV$^2$. It is worth noting that the mass exhibits a stability point around
$s_0 \sim$ 2.1 GeV$^2$. However, the Borel window lies above this point; it is only valid if $s_0 \geq s_0^{\rm min} = 5.3$~GeV$^2$. So we choose $s_0$ sightly large than this value.  The mass and coupling constant are calculated to be
\begin{eqnarray}
M_{3/2^-} &=& 2.05^{+0.09}_{-0.10}{\rm~GeV} \, ,
\label{eq:mass3-}
\\ \nonumber
f_{3/2^-} &=& 0.037^{+0.007}_{-0.007}{\rm~GeV^3} \, .
\label{eq:decay3-}
\end{eqnarray}

For the $J^P = 3/2^+$ state, we determine that its working regions are $2.09$~GeV$^2 \leq M_B^2 \leq 2.30$~GeV$^2$ and $10.5$~GeV$^2 \leq s_0 \leq 11.5$~GeV$^2$, and calculate its mass and coupling constant as
\begin{eqnarray}
M_{3/2^+} &=& 3.13^{+0.27}_{-0.18}{\rm~GeV} \, ,
\label{eq:mass3+}
\\ \nonumber
f_{3/2^+} &=& 0.074^{+0.015}_{-0.009}{\rm~GeV^3} \, .
\label{eq:decay3+}
\end{eqnarray}
The mass of the positive parity state is estimated to be approximately 1 GeV larger than that of the negative parity state with greater uncertainties (a trend that also applies to the coupling constant). This indicates that the current with a derivative couples more effectively to the negative parity state than to the positive parity state.

We next perform the same numerical analysis using the current $J$ in Eq.~(\ref{def:current1}) to investigate the $J^P=1/2^\pm$ states. To do a complete analyses, we have also performed an investigation by using a current without derivative~\cite{Chen:2012ex}:
\begin{eqnarray}
J^\prime_{\mu} = -\sqrt{3} \epsilon^{abc} s^T_aC\gamma_\mu s_b s_c \, ,
\label{current3}
\end{eqnarray}
whose spin parity is $3/2^+$. The results are summarised in Table~\ref{tab:results}.

\begin{table}[hpt]
\begin{center}
\renewcommand{\arraystretch}{1.4}
\caption{Masses and coupling constants extracted from the currents $J$ in Eq.~(\ref{def:current1}), $J_{\mu}$ in Eq.~(\ref{def:current2}), and $J^\prime_\mu$ in Eq.(~\ref{current3}). CC represents the coupling constant}
\begin{tabular}{c|c|c|c}
\hline\hline
{Current}&{state}&Mass[GeV]&CC[GeV$^3$]
\\ \hline\hline
$J$&$|\Omega;1/2^+\rangle$&$3.05^{+0.21}_{-0.15}$&$0.168^{+0.045}_{-0.040}$
\\
&$|\Omega;1/2^-\rangle$&$2.07^{+0.07}_{-0.07}$&$0.079^{+0.011}_{-0.011}$
\\
$J_\mu$&$|\Omega;3/2^+\rangle$&$3.13^{+0.27}_{-0.18}$&$0.074^{+0.015}_{-0.009}$
\\
&$|\Omega;3/2^-\rangle$&$2.05^{+0.09}_{-0.10}$&$0.037^{+0.007}_{-0.007}$
\\
$J^\prime_\mu$&$|\Omega^\prime;3/2^+\rangle$&$1.59^{+0.10}_{-0.12}$&$0.033^{+0.006}_{-0.006}$
\\
&$|\Omega^\prime;3/2^-\rangle$ &$3.15^{+0.16}_{-0.17}$&$0.092^{+0.018}_{-0.018}$
\\ \hline\hline
\end{tabular}
\label{tab:results}
\end{center}
\end{table}

\section{summary}

In this paper, we studied the recently observed $\Omega$ baryon $\Omega(2012)$ making use of QCD sum rules. We constructed the $P$-wave $\Omega$ baryon currents with a covariant derivative, whose spins are 1/2 and 3/2 by performing the proper spin projections. We then analyzed the parity-projected QCD sum rules to separate the contribution of the positive parity and negative parity states. Thus, we systematically studied in total four states with spin-parity $1/2^\pm$ and $3/2^\pm$, and applied the QCD sum rule method to calculate their masses and coupling constants.

The results are summarised in Table~\ref{tab:results}. We determined the mass of the $J^P = 1/2^-$ state as
\begin{eqnarray}
M_{1/2^-} &=& 2.07^{+0.07}_{-0.07}{\rm~GeV} \, ,
\label{eq:result_m_12}
\end{eqnarray}
and that of $J^P = 3/2^-$ as
\begin{eqnarray}
M_{3/2^-} &=& 2.05^{+0.09}_{-0.10}{\rm~GeV} \, .
\label{eq:result_m_32}
\end{eqnarray}
As both masses are consistent with the $\Omega(2012)$, it is likely that the $\Omega(2012)$ is a $P$-wave excited $\Omega$ baryon with three strange quarks. However, due to the closeness of Eqs.~(\ref{eq:result_m_12}) and (\ref{eq:result_m_32}), we cannot determine its spin quantum number in the present analysis.

We have only focused on studying the mass and coupling constant of the \(\Omega(2012)\) baryon thus far. Based on these results, we plan to explore its decay properties in the near future, as these are also important for understanding its internal structure. If its spin parity is \(J^P = 3/2^-\), it is likely to decay via \(S\)-wave into the final state \(\bar K \Xi(1530)\), although this decay will be suppressed due to a small phase space factor. It can also decay into the final \(\bar K \Xi\) state, but only as a \(D\)-wave final state, which would result in a small total decay width. On the other hand, if its spin parity is \(J^P = 1/2^-\), it would more easily decay into \(\bar K \Xi\) via \(S\)-wave, with no phase space suppression, while the decay to \(\bar K \Xi(1530)\) would proceed via \(D\)-wave, leading to a significantly larger total decay width. All these initial expectations need to be validated through an actual QCD sum rule analysis. Furthermore, a QCD sum rule analysis using a five-quark current corresponding to the molecular state \(\bar K \Xi(1530)^{\ast}\) will also need to be conducted in the future.

\section{Acknowledgments}

N.S. is supported by the China Scholarship Council under Grant No.~202306090272.
H.X.C. is supported by
the National Natural Science Foundation of China under Grant No.~12075019,
the Jiangsu Provincial Double-Innovation Program under Grant No.~JSSCRC2021488,
and
the Fundamental Research Funds for the Central Universities.
P.G. is supported by KAKENHI under Contract No. JP22H00122.
A.H. is supported in part by the Grants-in-Aid for Scientific Research [Grant No. 21H04478(A), 24K07050(C)].
\vfill\eject


\begin{thebibliography}{999}
\bibitem{Belle:2018mqs}
J.~Yelton \textit{et al.} [Belle],
Phys. Rev. Lett. \textbf{121}, no.5, 052003 (2018).
\bibitem{Belle:2021gtf}
Y.~Li \textit{et al.} [Belle],
Phys. Rev. D \textbf{104} (2021) no.5, 052005.

\bibitem{Belle:2022mrg}
 [Belle],
[arXiv:2207.03090 [hep-ex]].

\bibitem{Aliev:2018syi}
T.~M.~Aliev, K.~Azizi, Y.~Sarac and H.~Sundu,
Phys. Rev. D \textbf{98} (2018) no.1, 014031.

\bibitem{Aliev:2018yjo}
T.~M.~Aliev, K.~Azizi, Y.~Sarac and H.~Sundu,
Eur. Phys. J. C \textbf{78} (2018) no.11, 894.

\bibitem{Polyakov:2018mow}
M.~V.~Polyakov, H.~D.~Son, B.~D.~Sun and A.~Tandogan,
Phys. Lett. B \textbf{792} (2019), 315-319.

\bibitem{Wang:2018hmi}
Z.~Y.~Wang, L.~C.~Gui, Q.~F.~L\"u, L.~Y.~Xiao and X.~H.~Zhong,
Phys. Rev. D \textbf{98} (2018) no.11, 114023.

\bibitem{Xiao:2018pwe}
L.~Y.~Xiao and X.~H.~Zhong,
Phys. Rev. D \textbf{98} (2018) no.3, 034004.

\bibitem{Liu:2019wdr}
M.~S.~Liu, K.~L.~Wang, Q.~F.~L\"u and X.~H.~Zhong,
Phys. Rev. D \textbf{101} (2020) no.1, 016002.

\bibitem{Arifi:2022ntc}
A.~J.~Arifi, D.~Suenaga, A.~Hosaka and Y.~Oh,
Phys. Rev. D \textbf{105} (2022) no.9, 094006.

\bibitem{Menapara:2021vug}
C.~Menapara and A.~K.~Rai,
Chin. Phys. C \textbf{46} (2022) no.10, 103102.

\bibitem{Wang:2022zja}
K.~L.~Wang, Q.~F.~L\"u, J.~J.~Xie and X.~H.~Zhong,
Phys. Rev. D \textbf{107} (2023) no.3, 034015.

\bibitem{Zhong:2022cjx}
H.~H.~Zhong, R.~H.~Ni, M.~Y.~Chen, X.~H.~Zhong and J.~J.~Xie,
Chin. Phys. C \textbf{47} (2023) no.6, 063104.

\bibitem{Lin:2018nqd}
Y.~H.~Lin and B.~S.~Zou,
Phys. Rev. D \textbf{98} (2018) no.5, 056013.

\bibitem{Valderrama:2018bmv}
M.~P.~Valderrama,
Phys. Rev. D \textbf{98} (2018) no.5, 054009.

\bibitem{Pavao:2018xub}
R.~Pavao and E.~Oset,
Eur. Phys. J. C \textbf{78} (2018) no.10, 857.

\bibitem{Huang:2018wth}
Y.~Huang, M.~Z.~Liu, J.~X.~Lu, J.~J.~Xie and L.~S.~Geng,
Phys. Rev. D \textbf{98} (2018) no.7, 076012.

\bibitem{Gutsche:2019eoh}
T.~Gutsche and V.~E.~Lyubovitskij,
J. Phys. G \textbf{48} (2020) no.2, 025001.

\bibitem{Ikeno:2020vqv}
N.~Ikeno, G.~Toledo and E.~Oset,
Phys. Rev. D \textbf{101} (2020) no.9, 094016.

\bibitem{Zeng:2020och}
C.~H.~Zeng, J.~X.~Lu, E.~Wang, J.~J.~Xie and L.~S.~Geng,
Phys. Rev. D \textbf{102} (2020) no.7, 076009.

\bibitem{Lu:2020ste}
J.~X.~Lu, C.~H.~Zeng, E.~Wang, J.~J.~Xie and L.~S.~Geng,
Eur. Phys. J. C \textbf{80} (2020) no.5, 361.

\bibitem{Liu:2020yen}
X.~Liu, H.~Huang, J.~Ping and D.~Chen,
Phys. Rev. C \textbf{103} (2021) no.2, 025202.

\bibitem{Ikeno:2022jpe}
N.~Ikeno, W.~H.~Liang, G.~Toledo and E.~Oset,
Phys. Rev. D \textbf{106} (2022) no.3, 034022.

\bibitem{Hu:2022pae}
X.~Hu and J.~Ping,
Phys. Rev. D \textbf{106} (2022) no.5, 054028.

\bibitem{Lu:2022puv}
Q.~F.~L\"u, H.~Nagahiro and A.~Hosaka,
Phys. Rev. D \textbf{107} (2023) no.1, 014025.

\bibitem{Ioffe:1981kw}
B.~L.~Ioffe,
Nucl. Phys. B \textbf{188} (1981), 317-341
[erratum: Nucl. Phys. B \textbf{191} (1981), 591-592].

\bibitem{pdg}
P.~A.~Zyla \textit{et al.} [Particle Data Group],
PTEP \textbf{2020} (2020) no.8, 083C01.

\bibitem{Ovchinnikov:1988gk}
A.~A.~Ovchinnikov and A.~A.~Pivovarov,
Sov. J. Nucl. Phys. \textbf{48} (1988), 721-723

\bibitem{Yang:1993bp}
K.~C.~Yang, W.~Y.~P.~Hwang, E.~M.~Henley and L.~S.~Kisslinger,
Phys. Rev. D \textbf{47} (1993), 3001-3012.

\bibitem{Ellis:1996xc}
J.~R.~Ellis, E.~Gardi, M.~Karliner and M.~A.~Samuel,
Phys. Rev. D \textbf{54} (1996), 6986-6996.

\bibitem{Ioffe:2002be}
B.~L.~Ioffe and K.~N.~Zyablyuk,
Eur. Phys. J. C \textbf{27} (2003), 229-241.

\bibitem{Jamin:2002ev}
M.~Jamin,
Phys. Lett. B \textbf{538} (2002), 71-76.

\bibitem{Gimenez:2005nt}
V.~Gimenez, V.~Lubicz, F.~Mescia, V.~Porretti and J.~Reyes,
Eur. Phys. J. C \textbf{41} (2005), 535-544.

\bibitem{Narison:2011xe}
S.~Narison,
Phys. Lett. B \textbf{706} (2012), 412-422.

\bibitem{Narison:2018dcr}
S.~Narison,
Int. J. Mod. Phys. A \textbf{33} (2018) no.10, 1850045.

\bibitem{Chen:2012ex}
H.~X.~Chen,
Eur. Phys. J. C \textbf{72} (2012), 2180.
 \end{thebibliography}
\end{document}